\documentclass[final,3p,times,twocolumn]{elsarticle}

 \biboptions{comma,sort&compress}

\usepackage{here}
 \usepackage{graphicx}
  \usepackage{epsfig}
  \usepackage{amsmath}

\def\beq{\begin{equation}}
\def\eeq{\end{equation}}
\def\bea{\begin{eqnarray}}
\def\eea{\end{eqnarray}}

\journal{Nuc. Phys. (Proc. Suppl.)}

\begin{document}

\begin{frontmatter}

\title{The $a_1(1260)$ meson and chiral symmetry restoration and deconfinment at finite temperature QCD}

\author[ronde]{C. A. Dominguez\fnref{nrf}}
  \ead{Cesareo.Dominguez@uct.ac.za}

\author[ronde,puc]{M. Loewe\corref{speak}\fnref{conicyt,anillos}}
  \ead{mloewe@fis.puc.cl}

\author[ronde]{Y. Zhang}
  \ead{Yingwen.Zhang@uct.ac.za}

\address[ronde]{Centre for Theoretical \& Mathematical Physics,University of Cape Town, Rondebosch 7700, South Africa}

\address[puc]{Facultad de F\'{\i}sica, Pontificia Universidad Cat\'olica de Chile,
                    Casilla 306, Santiago 22, Chile.  }

\cortext[speak]{Speaker}

\fntext[nrf]{Supported by NRF South Africa}
\fntext[conicyt]{Supported by Fondecyt Chile under grant 1095217}
\fntext[anillos]{Supported by Proyecto Anillos ACT119}

\begin{abstract}
\noindent
We consider the light-quark axial-vector current correlator in the framework of
thermal QCD sum rules to: (a) find a relation between
chiral-symmetry restoration and deconfinement, and (b) determine the
temperature behaviour of the $a_1(1260)$ width and coupling. Our
results show that deconfinement takes place at a slightly lower
temperature than chiral-symmetry restoration.This difference is not
significant given the accuracy of the method. The behaviour of the
$a_1(1260)$ parameters is consistent with quark-gluon deconfinement,
since the width grows and the coupling decreases with increasing
temperature.
\end{abstract}

\begin{keyword}
Finite temperature field theory, hadron physics.

\end{keyword}

\end{frontmatter}

\section{Introduction}

A quark-gluon deconfinement  phenomenological order parameter was first
introduced in the vector channel, and in the framework of
thermal QCD sum rules in \cite{BS}. It is given by the squared energy threshold,
$s_0(T)$, for the onset of perturbative QCD (PQCD) in the hadronic
spectral function. In general, around $s_0(T=T_c)$, the resonances in
the spectrum, in any channel, will be either no longer present or
will become very broad. When the temperature approaches the critical
value for deconfinement, $T_c$, one would expect hadrons to
disappear from the spectral function, leading to a description based
entirely on  PQCD. It was shown in \cite{BS}  that $s_0(T)$ in the vector channel is a
decreasing function of the temperature, together with the coupling
of the $\rho$-meson to the vector current. This analysis, however,
was performed only in the zero-width approximation. Resonance
broadening was proposed in \cite{DL1}-\cite{DL2}, and subsequently
confirmed in other applications \cite{GAMMA3}, as an important
phenomenological information for the deconfinement transition.

\noindent
 QCD sum rules in the axial-vector channel, \cite{DL1},
established a  link between deconfinement and chiral-symmetry
restoration at finite temperature. This discussion was improved
later in \cite{GATTO1}, and recently updated and extended also to
finite density in \cite{DL3}. These analyses indicate that the
temperature at which $s_0(T)$ vanishes is very close to that at
which the quark condensate vanishes. The results suggest that both
phase transitions take place at roughly the same temperature. The
analyses of \cite{DL1}, \cite{GATTO1}-\cite{DL3} made use of the
finite energy QCD sum rule (FESR) of the lowest dimension ($d=2$) in
the axial-vector channel, assuming full saturation of the hadronic
spectral function by the pion pole. This assumption is not justified
if one were to consider the next two FESR of dimension $d=4$ and
$d=6$. In fact, already at $T=0$ one finds that the values of the
condensates of dimension $d=4$ and $d=6$ that follow from the second
and third FESR are barely consistent with results obtained from
experimental data \cite{DS1}. This suggests the presence of
additional hadronic contributions. In fact, the data in this channel
include also the $a_1(1260)$ resonance.

\noindent
 Using the first three FESR, we will reconsider the
light-quark axial-vector channel using a hadronic spectral function
involving the pion pole as well as the $a_1(1260)$ resonance. We
will obtain information on the temperature behavior of the $a_1(1260)$
coupling and hadronic width. The results show that $s_0(T)$ vanishes
at a critical temperature some 10\% below that for chiral-symmetry
restoration. This difference is not relevant within the accuracy of the method.
The $a_1(1260)$ coupling initially increases with increasing $T$ up
to $T/T_c \simeq 0.7$, and then decreases sharply up to $T_c$.
Finally, the hadronic width of the $a_1(1260)$ remains constant up
to $T/T_c \simeq 0.6$, increasing sharply thereafter. These
behaviours are consistent with a quark-gluon deconfinement scenario.

\section{FESR at ${\bf{T=0}}$}
\noindent We consider the correlator of light-quark axial-vector
currents
%Eq.1
\begin{eqnarray}
\Pi_{\mu\nu} (q^{2}) \!  &=& \! i  \int d^{4} x \; e^{i q x}  <0|T(
A_{\mu}(x) \;,  A_{\nu}^{\dagger}(0))|0> \nonumber \\ [.3cm] &=&
-g_{\mu\nu}\, \Pi_1(q^2) + q_\mu q_\nu\, \Pi_0(q^2)  \; ,
\end{eqnarray}
where $A_\mu(x) = : \bar{d}(x) \gamma_\mu \, \gamma_5 u(x):$ is the
(charged) axial-vector current, and $q_\mu = (\omega, \vec{q})$ is
the four-momentum carried by the current. Concentrating on e.g.
$\Pi_0(q^2)$ and invoking the Operator Product Expansion (OPE) of
current correlators at short distances, one has
%Eq.2
\begin{equation}
4 \pi^2\,\Pi_0(q^2)|_{\mbox{\scriptsize{QCD}}} = C_0 \, \hat{I} +
\sum_{N=1} \frac{C_{2N} (q^2,\mu^2)}{Q^{2N}} \langle
\hat{\mathcal{O}}_{2N} (\mu^2) \rangle \;, \label{OPE}
\end{equation}
where $Q^2 \equiv - q^2$, $\langle \hat{\mathcal{O}}_{2N} (\mu^2)
\rangle \equiv \langle0| \hat{\mathcal{O}}_{2N} (\mu^2)|0 \rangle$,
$\mu^2$ is a renormalization scale, the Wilson coefficients $C_N$
depend on the Lorentz indexes and
 quantum numbers of the currents, and on the local gauge invariant operators ${\hat{\mathcal{O}}}_N$ built
  from the quark and gluon fields. The Wilson coefficients are calculable in PQCD. The unit
   operator above has dimension $d=0$ and $C_0 \hat{I}$
   stands for the purely perturbative contribution.
    The dimension $d=4$ term, a renormalization group invariant quantity, is given by
%Eq.3
\begin{equation}
C_4 \langle \hat{\mathcal{O}}_{4}  \rangle = \frac{\pi}{6} \langle
\alpha_s G^2\rangle + 2 \pi^2 (m_u + m_d) \langle\bar{q} q \rangle ,
\label{C4}
\end{equation}
where the second term is negligible in comparison with the gluon
condensate. The leading power correction of dimension $d=6$ is the
four-quark condensate. In the vacuum
 saturation approximation it becomes
%Eq.4
\begin{equation}
C_6 \langle \hat{\mathcal{O}}_{6}  \rangle = \frac{704}{81} \,\pi^3
\, \alpha_s \,|\langle \bar{q} q \rangle|^2\;, \label{C6}
\end{equation}
having a  mild dependence on the renormalization scale. This
approximation has no theoretical
 justification. Hence, there is no reliable way of estimating corrections, which
  appear to be rather large from comparisons between Eq. (\ref{C6}) and determinations from data \cite{DS1}.
  This poses no problem for the finite temperature analysis, where Eq.(\ref{C6}) is only used to normalize results at
   $T=0$, and one is interested in the behavior of ratios.
 Cauchy's theorem
in the complex squared energy $s$-plane, leads us
 to the FESR (at leading order in PQCD)
%Eq.5
\begin{eqnarray}
\!\!\!\!\!&\!\! \!\!\!\!\!\!\!\!\!\!\!(-)^{(N-1)}& \!\!\!\! C_{2N}
\langle {\mathcal{\hat{O}}}_{2N}\rangle = 4 \pi^2 \int_0^{s_0} ds\,
s^{N-1} \,\frac{1}{\pi} {\mbox{Im}}
 \Pi_0(s)|_{\mbox{\scriptsize {HAD}}} \nonumber \\
[.1cm] &-&  \frac{s_0^N}{N} \left[1+{\mathcal{O}}(\alpha_s)\right]
\; (N=1,2,\cdots) .\label{FESR}
\end{eqnarray}
%Fig.1
\begin{figure}[ht]
\includegraphics[scale=0.4
]{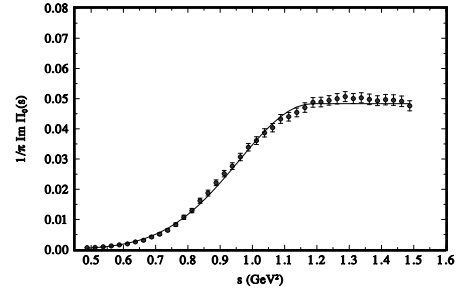} \caption{{\protect\small {The ALEPH data in the
axial-vector channel \cite{ALEPH}, and in the resonance region
together with our fit in the region relevant to the FESR.}}}
\label{figure1}
\end{figure}

The normalization of the correlator in PQCD is
%Eq.6
\begin{equation}
{\mbox{Im}}\, \Pi_0(s)|_{\mbox{\scriptsize {QCD}}} =
\frac{1}{4\,\pi} \left[1 + {\cal{O}}(\alpha_s(s))\right]\;.
\label{NORM}
\end{equation}
At $T=0$ the radiative corrections above are known up to five-loop
order, i.e. ${\cal{O}}(\alpha_s^4)$, in PQCD. Higher dimensional
condensates are poorly known \cite{DS1} and will not be considered
here.

%\begin{figure}[ht]
%\includegraphics[scale=0.4
%]{FPI_T.EPS} \caption{
%{\protect\small {The quark condensate $\langle
%\bar{q} q \rangle (T)/\langle \bar{q} q \rangle(0) =
%f_\pi^2(T)/f_\pi^2(0)$ as a function of $T/T_c$ in the chiral limit
%($m_q=M_\pi=0$) with $T_c = 197\; {\mbox{MeV}}$ \cite{QIN} (solid
%curve), and for finite quark masses from a fit to lattice QCD
%results \cite{LATTICE} (dotted curve).}}} \label{figure1}
%\end{figure}

In the hadronic sector the spectral function involves the pion pole
followed by the $a_1(1260)$ resonance
%Eq.7
\begin{equation}
{\mbox{Im}}\, \Pi_0(s)|_{\mbox{\scriptsize{HAD}}} = 2 \,\pi\,
f_\pi^2 \;\delta(s) + {\mbox{Im}}\, \Pi_0(s)|_{a_1}\;, \label{HAD}
\end{equation}
where $f_\pi = 92.21\, \pm\, 0.14 \; {\mbox{MeV}}$ \cite{PDG} is the
pion decay constant and the pion mass has been neglected. We have
made a fit to the ALEPH data \cite{ALEPH} (at T=0); for details, see
 \cite{paperoriginal}. This fit together with
the ALEPH data is shown in Fig.1 up to $s \simeq 1.5\,
{\mbox{GeV}}^2$ (the FESR determine $s_0 = 1.44\, {\mbox{GeV}}^2$).
The pion decay constant is related to the quark condensate through
the Gell-Mann-Oakes-Renner relation
%Eq.9
\begin{equation}
2\,f_\pi^2\,M_\pi^2 = - (m_u + m_d)\langle 0| \bar{u} u + \bar{d}
d|0\rangle\;. \label{GMOR}
\end{equation}
Chiral corrections to this relation are at the 5\% level
\cite{GMOR}. At finite temperature deviations are
 negligible except very close to the critical temperature \cite{GMORT}.\\
The first three FESR can now be used to determine the PQCD threshold
$s_0$, and the $d=4$ and $d=6$ condensates. These results will be
used to normalize all finite temperature results. The value of $s_0$
obtained by saturating the hadronic spectral function with only the
pion pole, and to leading order in PQCD, is $s_0 \simeq 0.7\,
{\mbox{GeV}}^2$, as in \cite{DL1}, \cite{GATTO1}-\cite{DL3}. This
value increases  to $s_0 =1.15 \, {\mbox{GeV}}^2$ once the
$a_1(1260)$ contribution is taken into account, and becomes $s_0
=1.44 \, {\mbox{GeV}}^2$ with PQCD to five-loop order.

%Fig.3
%\begin{figure}[ht]
%\includegraphics[scale=0.4]{C4O4.EPS}\caption{{\protect\small
%{The gluon condensate $C_4\langle {\cal{O}}_4\rangle(T)/C_4\langle
%{\cal{O}}_4\rangle(0) $ as a function of $T/T_c$  from lattice QCD
%results \cite{G2_Lattice}.  Solid squares and circles correspond to
%2 and 4 quark flavours, respectively, and error bars are the size of
%the points. The dotted curve is a fit to these data and the solid
%curve a smoothed fit.}}} \label{figure3}
%\end{figure}
\section{Finite Energy QCD Sum Rules at ${\bf{T\neq 0}}$}
\noindent
%Fig.2
\begin{figure}[hb]
\includegraphics[scale=0.4]{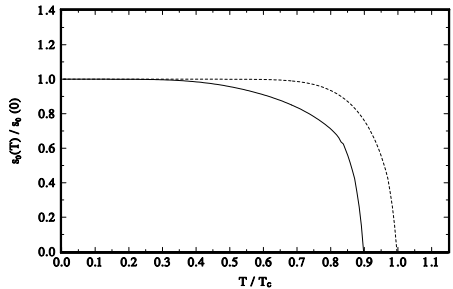}\caption{{\protect\small
{The continuum threshold $s_0(T)/s_0(0)$ signaling deconfinement
(solid curve)  as a function of $T/T_c$, together with
$f^2_\pi(T)/f^2_\pi(0) = \langle\bar{q} q\rangle(T)/\langle\bar{q}
q\rangle(0)$ signaling chiral-symmetry restoration (dotted
curve).}}} \label{figure4}
\end{figure}
The extension of the QCD sum rule method to finite temperature
implies a $T$-dependence in
 the OPE, as well as in the hadronic parameters. The Wilson coefficients
 and the vacuum condensates become temperature dependent. The strong
coupling $\alpha_s(Q^2,T)$ now
   depends on two scales, the ordinary QCD scale $\Lambda_{QCD}$ associated with the momentum transfer, and the critical
    temperature scale $T_c$ associated with temperature. This poses no problems in the asymptotic
     freedom region and at very high temperatures, $T >> T_c$,  where PQCD can be applied. However, the QCD
      sum rule method approaches $T_c$ from below, so that the presence of this second
       scale is problematic. No satisfactory solution to this problem exists, so that analyses must be carried
        out at leading order in PQCD.
At this order in PQCD there are two thermal corrections to
Eq.(\ref{NORM}). In the
 time-like region ($q^2 >0$), we have the  annihilation term which in the static limit (${\bf{q}} \rightarrow 0$) is
%Eq.10
\begin{equation}
{\mbox{Im}} \, \Pi_0^{+}(\omega,T) = \frac{1}{4\,\pi} \left[1 - 2 \,
n_F\left(\frac{\omega}{2 T}\right)\right]\;.
\end{equation}
In the space-like region we have the scattering term, which is
associated to
 a cut centered at the origin on the real axis in the complex energy $\omega \equiv \sqrt{s}$-plane
   of width $- |{\bf{q}}| \leq \omega \leq |{\bf{q}}|$ \cite{BS}. In the static limit  this  is given by
%Eq.11
\begin{eqnarray}
{\mbox{Im}} \, \Pi_0^{-}(\omega,T) &=& \frac{4}{\pi} \;
\delta(\omega^2) \; \int_0^\infty \, y \; n_F\left(
\frac{y}{T}\right) \; dy \nonumber \\ [.3cm] &=&\frac{\pi}{3}\; T^2
\, \delta(\omega^2)\;, \label{ST}
\end{eqnarray}
where $n_F(z) = 1/(1 + e^{-z})$ is the Fermi thermal function, and
the chiral limit was assumed.
%%%%%%%%%%%%%%%%%%%%%%%%%%
These perturbative results are valid for $T\geq 0$ at the one-loop
level in QCD, so that temperature effects
 develop smoothly from their $T=0$ values. Non-perturbative contributions will be added later in the framework
  of the OPE.

%%%%%%%%%%%%%%%%%%%%%%%%%%

In the hadronic sector and at finite temperature, masses, couplings,
and widths become $T$-dependent. Hadronically stable particles, e.g.
the pion, with $\Gamma(0) = 0$ develop a width. The important
parameters signaling deconfinement
  are the hadronic width and coupling, but not the mass. A vanishing mass at $T=T_c$
    would not signal deconfinement, unless the width diverges at such a temperature. But then the value of the mass
      becomes irrelevant. This is actually what
         QCD sum rule analyses show \cite{GAMMA3}. A notable exception are the scalar, pseudoscalar,
         and vector charm-anti-charm
          states which survive beyond $T_c$ \cite{GAMMA2}.

%Fig.3
\begin{figure}[hb]
\includegraphics[scale=0.4]{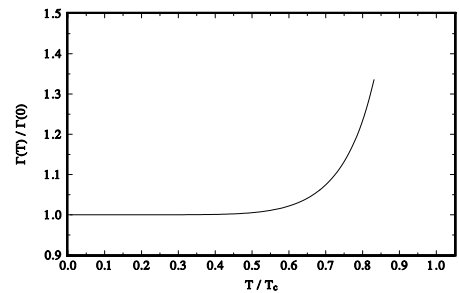}\caption{{\protect\small
{The hadronic width of the $a_1(1260)$ resonance
$\Gamma_{a_1}(T)/\Gamma_{a_1}(0)$  as a function of $T/T_c$.}}}
\label{figure5}
\end{figure}
The temperature behavior of the quark condensate, equivalently
$f_\pi^2$, was obtained, in the chiral limit,
 in the framework of the Schwinger-Dyson equation \cite{QIN}, and for
  finite quark masses from a fit to lattice QCD results \cite{DL3},
  \cite{LATTICE}. See \cite{paperoriginal} for details.
  The critical temperature is $T_c = 197 \; {\mbox{MeV}}$. In
   this temperature region the quark condensate in the chiral limit is essentially
   the same as that for finite quark masses.
The gluon condensate was obtained from a fit to  lattice QCD
determinations \cite{G2_Lattice}, adjusted to $T_c = 197\;
{\mbox{MeV}}$. The first three thermal FESR become
%Eq.12
\begin{eqnarray}
8  \pi^2 f^2_\pi(T) \!\!&=& \!\!\frac{4}{3}  \pi^2  T^2 \!\! +
\int_0^{s_0(T)}ds \,\left[1 - 2\, n_F \left(\frac{\sqrt{s}}{2 T}
\right) \right] \nonumber \\ [.1cm] &-& \!\! 4 \,\pi^2\,
\int_0^{s_0(T)} ds\, \frac{1}{\pi}\, {\mbox{Im}}\, \Pi_0(s,T)|_{a_1}
\;, \label{FESRT1}
\end{eqnarray}
%Eq.13
\begin{eqnarray}
- C_{4}\langle {\mathcal{\hat{O}}}_{4}\rangle(T)\!\! &=& \!\!4 \pi^2
\int_0^{s_0(T)} ds\, s \frac{1}{\pi} {\mbox{Im}}\, \Pi_0(s)|_{a_1}
\nonumber \\ [.1cm] &-&  \int_0^{s_0(T)}ds \, s \!\left[1 - 2
n_F\left(\frac{\sqrt{s}}{2 T}\right)\right] ,\label{FESRT2}
\end{eqnarray}
%Eq.14
\begin{eqnarray}
&& \!\!\!\!\!\!\!\!\!\!\!C_{6}\langle
{\mathcal{\hat{O}}}_{6}\rangle(T) \,\, = \,\, 4 \pi^2
\int_0^{s_0(T)}
ds\, s^2 \frac{1}{\pi} {\mbox{Im}}\, \Pi_0(s)|_{a_1} \nonumber \\
[.1cm] &-& \int_0^{s_0(T)}ds \; s^2 \left[1 - 2
n_F\left(\frac{\sqrt{s}}{2 T}\right)\right] \;.\label{FESRT3}
\end{eqnarray}
These equations determine the continuum threshold $s_0(T)$, the
coupling of the $a_1(1260)$ to the
 axial-vector current, $f_{a_1}(T)$, and its width $\Gamma_{a_1}(T)$, using as input the thermal quark condensate
  (or $f^2_\pi(T)$), the thermal $d=4,6$ condensates, and assuming the $a_1(1260)$ mass to be temperature independent, as
   supported by results in many channels \cite{GAMMA3}, \cite{GAMMA2},
   \cite{NUCLEON}.
\section{Results and Conclusions}
The FESR have solutions for the three parameters, $s_0(T)$,
$f_{a_1}(T)$, and $\Gamma_{a_1}(T)$, up to $T \simeq (0.85-0.90)
\,T_c$, a temperature at which $s_0(T)$ reaches
 its minimum. An inspection of Eq.(\ref{FESRT1}) shows that
disregarding the $a_1(1260)$ contribution, $s_0(T)$ would vanish at
 a lower critical temperature than $f_\pi(T)$ (or $\langle\bar{q} q\rangle(T)$). Making the
  rough approximation of neglecting the thermal factor $n_F(\sqrt{s}/2T)$ in the second term on
   the r.h.s. of Eq.(\ref{FESRT1}) leads to $s_0(T) \simeq 8 \, \pi^2\, f^2_\pi(T) - (4/3)\, \pi^2 \, T^2$.
%Fig.4
\begin{figure}[ht]
\includegraphics[scale=0.4]{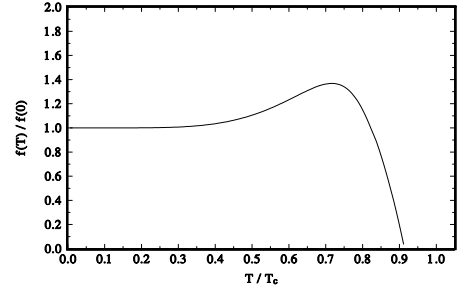}\caption{{\protect\small
{The coupling of the $a_1(1260)$ resonance $f_{a_1}(T)/f_{a_1}(0)$
as a function of $T/T_c$.}}} \label{figure6}
\end{figure}
This feature remains valid even after including the $a_1(1260)$ in
the FESR, as shown in Fig.\ref{figure4}, corresponding to the
solution for $s_0(T)$ using all three FESR. This 10\% difference is
well
 within the accuracy of the method. The behavior of the width is shown in Fig.\ref{figure5}, and that of the
  coupling in Fig.\ref{figure6}. The rise of the width, and the fall of the coupling are indicative of a transition
   to a quark-gluon deconfined phase at $T = T_c$.
%%%%%%%%%%%%%%%%%%%%%%%%%%%%%%%%%%%%%%%%%%%%%%%%%%%%%%%%%%%%%%%%%%%%%%%%%%%%%%%%

\section{Acknowledgments}
This work has been supported in part by NRF (South Africa), FONDECYT
1095217, 1120770 (Chile), and Proyecto Anillos ACT 119 (Chile).

\end{document}